\documentstyle[aps,prb,floats,epsfig]{revtex}

\title{Observation of the screening signature in the lateral photovoltage of electrons in the Quantum Hall regime}
\author{H. van Zalinge$^{1}$, B. \"{O}zyilmaz$^{2}$\cite{present1}, A. B\"{o}hm$^{2}$\cite{present2}, R.W. van der Heijden$^{1}$, J.H. Wolter$^{1}$, and P. Wyder$^{2}$}
\address{$^{1}$COBRA Inter-University Research Institute, Department of Physics,
Eindhoven
University of Technology, P.O. Box 513, NL-5600
MB Eindhoven, The Netherlands}
\address{$^{2}$Hochfeld-Magnetlabor, Max-Planck Institut f\"{u}r Festk\"{o}rperforschung, and Centre National de la Recherche Scientifique, B.P. 166, 25 Avenue des Martyrs, F-38042 Grenoble Cedex 9, France}

\begin{document}
\draft

\twocolumn[\hsize\textwidth\columnwidth\hsize\csname@twocolumnfalse\endcsname

\maketitle

\begin{abstract}
The lateral photovoltage generated in the plane of a two-dimensional electron system (2DES) by a focused light spot, exhibits a fine-structure in the quantum oscillations in a magnetic field near the Quantum Hall conductivity minima. A double peak structure occurs near the minima of the longitudinal conductivity oscillations. This is the characteristic signature of the interplay between screening and Landau quantization. \end{abstract}


\vskip2pc]

\narrowtext

\newcommand{\av}[1]{\mbox{$\langle #1 \rangle$}}

One of the most intriguing aspects of a two-dimensional electron system (2DES) in high magnetic fields is the distribution of electrons over compressible and incompressible states. The distinguishing property is the ability to screen external electric fields or charges. Screening requires a redistribution of charge and is therefore only possible in compressible states. Of particular interest is the spatial distribution of these states over the sample area and their variation with magnetic field. These phenomena are also at the basis of the Quantum Hall Effects (QHE)\cite{sarma}, where the interior of the sample is incompressible and the boundary consists of an alternating structure of compressible and incompressible regions in a self-consistent manner\cite{chklovskii}. These edge states have played a crucial role in theories of the QHE\cite{buttiker}.
 
Over the years, many experiments have been performed to investigate properties related to the compressibility, both by using sample-averaging probes as well as by using spatially resolved scanning techniques. One of the oldest methods is the capacitance technique\cite{capacitance}, which measures directly the compressibility through the screening. Related is a technique that measures in an inductive manner the current induced by the penetrating electric field from a backgate\cite{yahel}. Particularly interesting is the use of a scanning single electron transistor (SET) in the transparency mode\cite{zhitenev}. In addition, scanning probe methods exist that detect the electric potentials resulting from charging of the compressible regions through an external source, including electro-optic\cite{fontein}, capacitance\cite{tessmer}, atomic force microscopy\cite{cormick} and SET techniques\cite{zhitenev,wei}.

The most direct and elementary investigation of screening would be to see how the system responds to a local charge injection. Such an experiment is closely realized in the present work, where charge is injected by means of the internal photoelectric effect by a focused light spot, while the resulting {\em lateral} photovoltage in the plane of the 2DES is measured. Previously, this experiment was used to probe the edge channel structure under Quantum Hall (QH) conditions\cite{haren,shaskin,boehm}. In the present work, the unexpected finding is reported that the photovoltage has a strongly oscillatory variation with magnetic field in a narrow field range centered around a QH conductivity minimum. The oscillatory behaviour is analogous to that known from a variety of other experiments and is the characteristic signature of screening in competition with density of states effects. Therefore the present observation is attributed to the manifestation of screening effects in the lateral photovoltage, in conjunction with the transport of the excess charge.

Data are reported from two different GaAs-Al$_{0.33}$Ga$_{0.67}$As wafers: wafer 1 had an intermediate 2DES mobility $\mu \sim$ 50 m$^{2}$/Vs and a 2DES density $n \sim$ 5$\times$10$^{15}$ m$^{-2}$; wafer 2 had a very high 2DES mobility $\mu \sim$ 300 m$^{2}$/Vs and a 2DES density $n \sim$ 1$\times$10$^{15}$ m$^{-2}$. Two independent setups were used in which a small laser spot was scanned across the sample (see Fig 1a). One employed a cryogenic scanning device with a spotsize of about 5 $\mu$m\cite{heil}. An optical cryostat with a room temperature scanning system and a spotsize of about 25 $\mu$m is used in the other one\cite{haren}. The pumped liquid helium temperatures in both setups are $T \sim$ 1.1 K and $T \sim$ 1.4 K respectively. 

\begin{figure}[!t]
\begin{center}
\leavevmode
\epsfig{figure=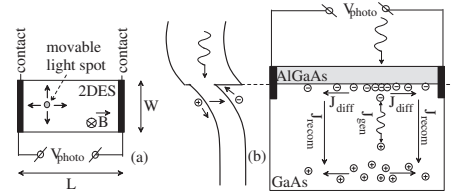,height=2.6cm,angle=0}
\end{center}
\caption{(a) Lay-out of sample with contacts and light spot. $L$ is always 4 mm, $W$ is either 2 or 0.5 mm (b) Schematic of heterojunction and sketch of internal generation, diffusion and recombination currents. Note that none of the dimensions is on scale. The cap layer is omitted.}
\label{fig1}
\end{figure}

The photon energy ($\lambda = 630$ nm) is above the bandgap of GaAs, so that electron-hole pairs are generated in the bulk of the GaAs. A very  thin (17 nm for wafer 1) and therefore nearly transparant GaAs cap layer is not expected to influence the experiments. The internal band bending near the GaAs-AlGaAs interface leads to separation of the electrons and holes (Fig. 1b). The electrons accumulate at the heterojunction in the 2DES whereas the holes escape into the bulk of the GaAs. The {\em lateral} voltage between Ohmic contacts to the 2DES is measured. The power incident at the sample is in the order of 1 $\mu$W. The light is modulated at a frequency near 70 Hz to enable the use of lock-in detection techniques. All experiments were done in the open circuit configuration ("voltage mode"), but lateral currents will still flow in the region where recombination processes occur ($J_{diff}$ and $J_{recom}$ in Fig. 1b)\cite{fontein2}.

The lateral photovoltage exhibits quantum oscillations with magnetic field, that were first investigated by B\"{o}hm {\it et al.} and termed LISHO (Light-Induced-Shubnikov-de Haas-Oscillations)\cite{boehm}. The striking new feature reported here is a double peak structure of the LISHO that occurs near sufficiently small, integer occupation factors $\nu$. In Fig. 2b it is shown for $\nu$=6 for several positions of the laserspot on a sample from wafer 1, along with the longitudinal conductivity track $\sigma_{xx}$($B$) in Fig. 2a. Fig. 2c displays the same feature at $\nu$=1 for a sample from the very different wafer 2. 

\begin{figure}[!t]
\begin{center}
\leavevmode
\epsfig{figure=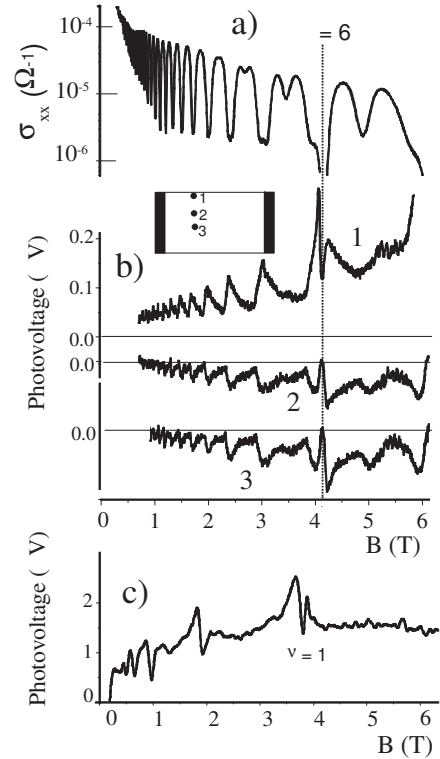,height=10cm,angle=0}
\end{center}
\caption{a) Wafer 1, $W$=2 mm, $\sigma_{xx}$($B$). b) The photovoltage as a function of magnetic field for three laserspot positions (see inset with sample). Note the sign change that occurs between track 1 and 2. These data were taken in a session with insufficient magnet current available to reach $\nu$=4 at 6.35 T. c) Wafer 2, $W$=0.5 mm. The photovoltage as a function of magnetic field for a laserspot near the edge of the sample.}
\label{fig2}
\end{figure}

A double peak near a QH-minimum results from a nonmonotonic dependence of a quantity on $\sigma_{xx}$. It is a general result from a competition between screening and localization effects and is known from many greatly varying experiments. The clearest example probably is the capacitance configuration, where the double peak occurs in the dissipation factor\cite{goodall}. For large conductivity, the dissipation is low because the field from the coupling capacitor is screened. At low conductivity, there is field in the 2DES, but the current is small. A second example is found in the attenuation of surface acoustic waves by the 2DEG\cite{wixforth}. The double-peak structure is also present in the acousto-electric effect\cite{esslinger1}, which is a SAW-induce DC voltage resulting from nonlinear interaction effects.

The previous two examples involve a time scale for charge relaxation, whereas the present experiments are essentially DC. A double peak for time-independent transport was obtained in the theoretical result for the drag transresistance of two closely spaced 2DES's\cite{bonsager}. The transresistance is the ratio of the voltage induced in one layer to the current generated in the other. Most remarkable in this theoretical result is that at the temperature for the present experiments (1.5 K), a double peak structure is obtained which is almost entirely confined to the field range where the density of states is in a minimum. This corresponds closely to the experiments of Fig. 2. A report of the observation of the predicted double peak in drag exists\cite{rubel}, but more recent work showed it to have an entirely different, spin-related origin\cite{lok}. The spin splitting is clearly visible in the conductivity trace of Fig. 2a (dips near 3.5 and 5 T). From the LISHO, particularly at position 2 and 3, it is clear that the spin splitting at 3.5 and 5 T is very different from the screening-induced splitting. The structure in the photovoltage near $\nu =6$ is therefore certainly not related to spin splitting. In addition, the structure for wafer 2 in Fig. 2c occurs at the $\nu =1$ spin-minimum. In contrast to the early drag-experiments\cite{rubel,hill}, the problems of distinguishing spin splitting from screening-induced splitting do not exist for the interpretation of the photovoltage data.  

For higher occupation factors (low $B$), the dip at the $\sigma_{xx}$-minimum is not resolved. The asymmetric, reverse-sawtooth-like lineshape of the oscillations in Fig. 2b is to be noted. It shows that the signal is not related in a straightforward way to 1/$\sigma_{xx}$\cite{haren,fontein2}, $\sigma_{xx}$\cite{shaskin}, $\rho_{xx}$ or $\rho_{xy}$. The magnetic field dependence at low $B$ resembles the variation of the carrier density in the upper-most occupied Landaulevel. It is known theoretically\cite{efros} that the screening properties of a 2DES are determined by this upper-most Landau level in a complicated and nonlinear way. The LISHO-lineshape at low $B$ could therefore reflect properties of the screening. 

\begin{figure}[!t]
\begin{center}
\leavevmode
\epsfig{figure=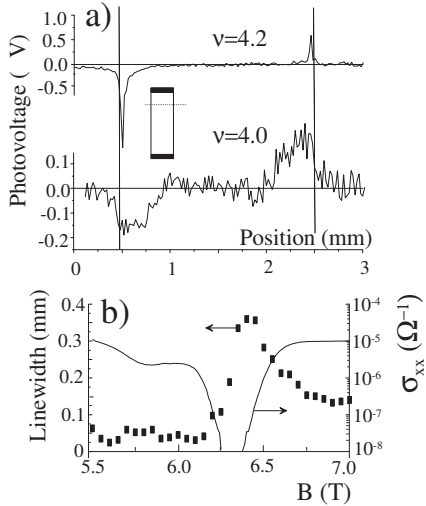,height=6.8cm,angle=0}
\end{center}
\caption{Data from wafer 1. a) Photovoltage linescans across sample (see inset) for occupation factors $\nu$=4.2 (6.05 T) and $\nu$=4.0 (6.35 T) respectively. b) Compilation of linewidths (full-width-at-half-maximum, FWHM) from data as in a) as a function of magnetic field together with $\sigma_{xx}$($B$).}
\label{fig3}
\end{figure}

Fig. 3 shows how the photovoltage varies with the position of the spot on the sample for different magnetic fields. The signal becomes large close to the edge, but, as follows from Fig. 2(b) and 2(c), near the edge the oscillating part is fairly small compared to the total signal. In the interior of the sample, away from the edges, the signal magnitude is small, but it varies strongly with magnetic field between zero and finite values. This behavior was most severe for wafer 2 (Fig. 2c), where it was actually not possible to obtain a signal in the bulk. The main phenomenon of the present work, the double peak, is observed for spot positions far from the edges. The signal {\it gradually} increases when approaching the sample edges from the sample side, and jumps quite abruptly to near zero when the sample edge is crossed. This can be seen even for the extreme case with the narrowest line as shown in Fig. 3(a). Both observations prove that the oscillatory phenomena presently reported are not an exclusive edge phenomenon, e.g. directly induced by the edge channel structure. A fortiori, they show that we are not dealing with spurious signals, generated by the laser spot when hitting the side of the mesa. The linewidths of linescans as in Fig. 3(a) (FWHM, averaged over the peaks at the two edges) are plotted as a function of $B$ in Fig. 3(b). The first observation is that the lines are narrow away from the QH regions and broad at and near them. Qualitatively, this is similar to the Hall potential profiles in current carrying Hall bars as found from Atomic Force Microscope (AFM) measurements\cite{ahlswede}. The profile was explained in terms of a decoupling of edge and bulk by an incompressible strip, causing the Hall potential to drop mainly near the edges outside the QH-regions. The length scales in the present case are several orders of magnitude larger than in the AFM experiments for both sample and probe, so structure due to (in)compressible strips will not be resolved. The variation with magnetic field of the linewidths in Fig. 3 demonstrates that the photovoltage is adversely affected by the macroscopic screening ability of the 2DES.

A second, very surprising result of Fig. 3(b) is the asymmetric variation of the linewidth when crossing a QH $\sigma_{xx}$-minimum. The linewidth variation with $B$ (Fig. 3b) resembles the inverse sawtooth-like $B$-dependence of the photovoltage at high $\nu$ (Fig. 2(b)). A similar variation was observed around the $\nu=6$ structure of Fig. 2b, though with less resolution. The linewidth is not simply a function of $\sigma_{xx}$, but depends also explicitly on the position of the Fermi level with respect to the center of the nearest Landau level. Again, this linewidth variation could reflect the particularities of the screening. The AFM-data\cite{ahlswede} also showed asymmetry with respect to the QH-mimima, however in a reverse manner as in Fig. 3(b). 

The magnetic field dependence of the data shown in Figs. 2 and 3 shows that the photovoltage is dominated by the 2DES. There is no direct need to invoke a role for the photogenerated holes in the substrate. As far as transport properties are concerned, this should be expected as the constraints to the motion in a strong magnetic field are most severe for high-mobility carriers ($\sigma_{xx} \rightarrow 0$ both classically and quantum mechanically). The screening is a property of the 2DES only, irrespective of what is screened. 

\begin{figure}[!b]
\begin{center}
\leavevmode
\epsfig{figure=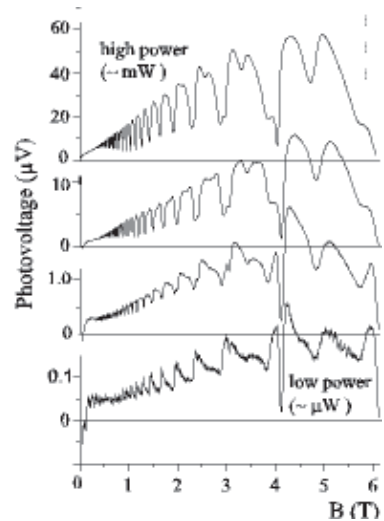,height=6.8cm,angle=0}
\end{center}
\caption{Photovoltage oscillations, same sample as Fig. 2a,b, for 4 different laserpowers varying approximately between 1$\mu$W and 1 mW.}
\label{fig4}
\end{figure}

Previously, quantum oscillations were reported in the lateral photovoltage for nonresonant far-infrared (FIR) absorption by the 2DES\cite{lorke}. They were attributed to direct thermoelectric effects due to the radiation-induced lattice heating. To identify possible heating effects in the present experiments, data were taken as a function of power level. As seen from Fig. 4 strong nonlinearities occur and at high power level the oscillations clearly are of entirely different origin than at low power. At the highest power levels, with photovoltages of similar magnitude as in Ref. \cite{lorke}, the double peak structure is completely washed out, the asymmetries disappear, and the photovoltage oscillations closely follow the $\rho_{xx}$ oscillations. The signal could well correspond to the longitudinal thermopower component $S_{xx}$, which has a shape resembling $\rho_{xx}$($B$)\cite{fletcher}. In contrast, the low power oscillations have a shape not corresponding to either one of the thermopower components. In particular, both $S_{xx}$ and $S_{xy}$ tend to zero near integer occupation, exactly opposite to the main trend in the present low-power data. 

In conclusion, a new magneto-oscillatory phenomenon has been observed in the lateral photoelectric effect, which is a novel probe for the quantum magnetotransport properties of a 2DES. The most pertinent effect is the double peak structure near a quantum Hall conductivity minimum, as a result of competition effects between screening and conductivity. 
With this method, it will be possible to study screening properties in a spatially resolved manner under varying conditions such as externally controlled screening by a parallel layer, the level of disorder, the presence of externally imposed currents, etcetera. In principle, there are no limitations for improving the resolution to the order of 1 $\mu$m or even below. 

We thank W.C. van der Vleuten and P.A.M. Nouwens for the growth and preparations of wafer 1, and C.T. Foxon and L.W. Molenkamp for providing wafer 2. B. \"{O}. and A.B. thank J. Riess and W. Dietsche for very fruitful discussions.


\begin{references}
\bibitem[*]{present1}Present Address: Department of Physics, New York University, 4 Washington Place, New York, New York 10003.
\bibitem[**]{present2}Present Address: Infineon Technologies AG, PO Box 800949, D-81609 Munich, Germany.
\bibitem{sarma}For an overview see: {\it Perspectives in Quantum Hall Effects}, edited by S. Das Sarma and A. Pinczuk (Wiley, New York, 1997).
\bibitem{chklovskii}D.B. Chklovskii, B.I. Shklovskii, and L.I. Glazman, Phys. Rev. B {\bf 46}, 4026 (1992).
\bibitem{buttiker}For an overview, see: M. B\"{u}ttiker, in {\it Semiconductors and Semimetals}, edited by M. Reed (Academic Press, San Diego, 1992), Vol. 44, p191.
\bibitem{capacitance}T.P. Smith, B.B. Goldberg, P.J. Stiles, and M. Heiblum, Phys. Rev. B {\bf 32}, 2696 (1985); J.P. Eisenstein, L.N. Pfeiffer, and K.W. West, Phys. Rev. Lett. {\bf 68}, 674 (1992); W. Chen, T.P. Smith III, M. B\"{u}ttiker, and M. Shayegan, Phys. Rev. Lett. {\bf 73}, 146 (1994); S. Takaoka, K. Oto, H. Kurimoto, K. Murase, K. Gamo, and S. Nishi, Phys. Rev. Lett. {\bf 72}, 3080 (1994); S. Takaoka, K. Oto, S. Uno, K. Murase, F. Nihey, and K. Nakamura, Phys. Rev. Lett. {\bf 81}, 4700 (1998); J.J. Mare\v{s}, J. Kri\v{s}tofik, and P. Hub\'{i}k, Phys. Rev. Lett. {\bf 82}, 4699 (1999).
\bibitem{yahel}E. Yahel, D. Orgad, A. Palevski, and H. Shtrikman, Phys. Rev. Lett. {\bf 76}, 2149 (1996); E. Yahel, A. Tsukernik, A. Palevski, and H. Shtrikman, Phys. Rev. Lett. {\bf 81}, 5201 (1998).
\bibitem{zhitenev}N.B. Zhitenev, T.A. Fulton, A. Yacoby, H.F. Hess, L.N. Pfeiffer, and K.W. West, Nature {\bf 404}, 473 (2000).
\bibitem{fontein}P.F. Fontein, J.A. Kleinen, P. Hendriks, F.A.P. Blom, J.H. Wolter, H.G.M. Lochs, F.A.J.M. Driessen, L.J. Giling, and C.W.J. Beenakker, Phys. Rev. B {\bf 43}, 12090 (1991).
\bibitem{tessmer}S.H. Tessmer, P.I. Glicofridis, R.C. Ashoori, L.S. Levitov, and M.R. Melloch, Nature {\bf 392}, 51 (1998).
\bibitem{cormick}K.L. McCormick, M.T. Woodside, M. Huang, M. Wu, P. McEuen, C. Duruoz, and J.S. Harris, Jr., Phys. Rev. B {\bf 59}, 4654 (1999).
\bibitem{wei}Y.Y. Wei, J. Weis, K. von Klitzing, and K. Eberl, Phys. Rev. Lett. {\bf 81}, 1674 (1998).
\bibitem{haren}R.J.F. van Haren, F.A.P. Blom, and J.H. Wolter, Phys. Rev. Lett. {\bf 74}, 1198 (1995); R.J.F. van Haren, W. de Lange, F.A.P. Blom, and J.H. Wolter, Phys. Rev. B {\bf 52}, 5760 (1995).
\bibitem{shaskin}A.A. Shaskin, A.J. Kent, J.R. Owers-Bradley, A.J. Cross, P. Hawker, and M. Henini, Phys. Rev. Lett. {\bf 79}, 5114 (1997).
\bibitem{boehm}A. B\"{o}hm, B. \"{O}zyilmaz, J. Heil, W.C. van der Vleuten, L.W. Molenkamp, U. Beyer, and P. Wyder, Inst. Phys. Conf. Ser. {\bf 164}, 231 (1999); A. B\"{o}hm, B. \"{O}zyilmaz, J. Heil, U. Beyer, P. Wyder, and J.H. Wolter, {\it unpublished}
\bibitem{heil}J. Heil, A. B\"{o}hm, M. Primke, P. Wyder, Rev. Sci. Instrum. {\bf 67}, 307 (1996).
\bibitem{fontein2}P.F. Fontein, P. Hendriks, J. Wolter, R. Peat, D.E. Williams, and J.-P. Andr\'{e}, J. Appl. Phys. {\bf 64}, 3085 (1988).
\bibitem{goodall}R.K. Goodall, R.J. Higgins, and J.P. Harrang, Phys. Rev. B {\bf 31}, 6597 (1985).
\bibitem{wixforth}A. Wixforth, J.P. Kotthaus, and G. Weimann, Phys. Rev. Lett. {\bf 56}, 2104 (1986); A. Wixforth, J. Scriba, M. Wassermeier, J.P. Kotthaus, G. Weimann, and W. Schlapp, Phys. Rev. B {\bf 40}, 7874 (1989).
\bibitem{esslinger1}A. Esslinger, A. Wixforth, R.W. Winkler, J.P. Kotthaus, H. Nickel, W. Schlapp, and R. L\"{o}sch, Solid State Commun. {\bf 84}, 939 (1992).
\bibitem{bonsager}M.C. B\o nsager, K. Flensberg, B. Y-K. Hu, A-P. Jauho, Phys. Rev. Lett. {\bf 77}, 1366 (1996).
\bibitem{rubel}H. Rubel, A. Fischer, W. Dietsche, K. von Klitzing, and K. Eberl, Phys. Rev. Lett. {\bf 78}, 1763 (1997).
\bibitem{lok}J.G.S. Lok, S. Kraus, M. Pohlt, W. Dietsche, K. von Klitzing, W. Wegscheider, and M. Bichler, Phys. Rev. B {\bf 63}, 041305(R) (2001).
\bibitem{hill}N.P.R. Hill, J.T. Nicholls, E.H. Linfield, M. Pepper, D.A. Ritchie, B. Y-K. Hu, and K. Flensberg, Physica B {\bf 249-251}, 868 (1998).
\bibitem{efros}A.L. Efros, Solid State Commun. {\bf 65}, 1281 (1988); A.L. Efros, Solid State Commun. {\bf 67}, 1019 (1988).
\bibitem{ahlswede}E. Ahlswede, P. Weitz, J. Weis, K. v. Klitzing, and K Eberl, Physica B {\bf 298}, 562 (2001).
\bibitem{lorke}A. Lorke, J.P. Kotthaus, J.H. English, and A.C. Gossard, Phys. Rev. B {\bf 53}, 1054 (1996).
\bibitem{fletcher}R. Fletcher, J.C. Maan, K. Ploog, and G. Weimann, Phys. Rev. B {\bf 33}, 7122 (1986).



\end{references}
\end{document}